\documentclass{article}
\usepackage{spconf,amsmath,graphicx,hyperref}
\usepackage{xcolor, tabularx, booktabs, makecell, siunitx, bm}
\usepackage{multirow, url}
\usepackage{subcaption}
\usepackage{balance}
\usepackage{stfloats}

\title{Exploring Speaker-Aware Engagement Modeling\\for Children with Autism Spectrum Disorder}
\name{
Kaleen Shrestha$^{1}$\qquad
Austin Joel Dympep$^{2}$\thanks{This is work done while Austin Joel Dympep was an intern at the University of Southern California through the IUSSTF-Viterbi program.} \qquad
Young Suh Hong$^{3}$ \qquad
Maja Matarić$^{1}$
}

\address{
$^{1}$ Department of Computer Science, Viterbi School of Engineering, University of Southern California \\
$^{2}$ Department of Computer Science and Engineering, National Institute of Technology Meghalaya \\
$^{3}$ Department of Systems and Information Engineering, University of Virginia
}
\begin{document}
\raggedbottom
%
\maketitle
\begin{abstract}
Prior work on engagement modeling in children with autism spectrum disorder (ASD) has demonstrated substantial individual variability in behavioral features, particularly in visual modalities. In this work, we extend the literature to acoustic features in naturalistic interactions, where multiple speakers and atypical interaction patterns complicate speaker-specific analysis. We investigated whether speaker diarization improved acoustic engagement modeling during month-long, in-home socially assistive robot (SAR) interventions with seven children with ASD. We trained classical machine learning models using acoustic features extracted from diarized child and parent/caregiver audio and compared their performance with models trained on undiarized audio. Diarized audio consistently yielded higher area under the receiver operating characteristic curve (AUROC) values, although these differences were not statistically significant. More importantly, the relative contributions of child and caregiver acoustic features varied across participants, indicating that the acoustic information associated with engagement was not consistent across children. These findings extend evidence of participant-level variability in ASD engagement modeling to the acoustic domain and demonstrate the feasibility of isolating speaker-specific vocalizations for studying engagement in complex, naturalistic interactions.


\end{abstract}

\section{Introduction}
\label{sec:intro}
Modeling engagement behavior of children with autism spectrum disorder (ASD) in in-home contexts is challenging because of variability in both naturalistic environments and behavioral expressions of engagement. Prior work has explored multimodal approaches to engagement modeling \cite{jain2020modeling, rudovic2018personalized}; however, in naturalistic, in-home settings children often interact with caregivers and others in the environment \cite{chorianopoulou2017engagement}, and such interactions influence engagement. Therefore, engagement modeling requires consideration of both child behavior and the surrounding social context \cite{kim2022parental}. 

Prior work on automated engagement modeling for children with ASD has highlighted substantial variability in behavioral expressions of engagement \cite{jain2020modeling, rudovic2018personalized}. Personalized and multimodal approaches have used visual, audio, and motion cues to estimate social engagement during child-robot interactions, with vocalizations providing complementary information \cite{javed2020automated, jain2020modeling, rudovic2018personalized}. However, naturalistic interactions often involve multiple speakers, complicating the attribution of acoustic information to the target individual \cite{rudovic2018personalized}. Speaker diarization has been used in ASD settings to distinguish child vocalizations from clinicians, caregivers, and other sources in noisy recordings \cite{gorodetski2019speaker}, demonstrating the feasibility of isolating child speech and motivating the use of speaker-specific acoustic features for engagement modeling.

Informed by prior work demonstrating substantial individual variability in ASD engagement modeling, we extend the state-of-the-art by focusing on acoustic engagement modeling in naturalistic interactions. We investigated the effect of speaker diarization on acoustic engagement modeling using data from a month-long, in-home socially assistive robot (SAR) intervention with seven children with ASD \cite{clabaugh2019long}. We trained multiple classical machine learning models using acoustic features extracted from diarized child and caregiver audio and compared their performance with models trained on features extracted from undiarized audio. Diarized audio consistently yielded higher area under the receiver operating characteristic curve (AUROC) values, although these differences were not statistically significant. We further examined the relative contributions of child and caregiver acoustic features to engagement classification and found that these contributions varied across participants. Together, these findings extend evidence of individual variability in ASD modeling to the acoustic domain and characterize how speaker-specific acoustic information contributes to engagement modeling in naturalistic interactions.

\section{Data}
\label{sec:data}
The data used in this work were collected during a month-long, in-home socially assistive robot (SAR) intervention with seven children with autism spectrum disorder (ASD), ages 3--7 years \cite{clabaugh2019long}. Participants interacted with a tabletop SAR and completed space-themed mathematics games on a touchscreen tablet while the robot provided feedback and encouragement verbally and through screen-based facial expressions and body movement. The collected data included webcam video and audio, and game-performance data, comprising approximately 21 hours of interactions across seven participants and multiple sessions. Study participants and their families determined session frequency; as a result, the dataset includes from as few as four sessions to as many as 15 sessions for a participant. We address issues of data and class imbalance during model training and evaluation (Section \ref{sec:methods-evaluation}). In all sessions, caregivers were present in the background.

The dataset had been human-annotated for binary engagement in prior work by Jain et al. \cite{jain2020modeling}. Each video frame was labeled as engaged (1) if the child participant was listening or responding to the SAR's prompts, interacting with the tablet, or seeking guidance from others in the room. All other behaviors were labeled as disengaged (0). We used these annotations in our modeling, described next.

\begin{table}[]
\caption{Number of audio clips for the different data configurations. Data between engaged and disengaged clips is slightly imbalanced, depending on the participant, and is more imbalanced for diarized audio, with engaged clips being more frequent than disengaged clips.}
\resizebox{\columnwidth}{!}{%
\begin{tabular}{l|cc|cccc}
\hline
\multicolumn{1}{c|}{\multirow{2}{*}{\textbf{Participant}}} & \multicolumn{2}{c|}{\textbf{Undiarized audio}} & \multicolumn{4}{c}{\textbf{Diarized audio}}                                                                                                                                                                                                                                                            \\
\multicolumn{1}{c|}{}                                      & \textbf{Engaged}     & \textbf{Disengaged}     & \textbf{\begin{tabular}[c]{@{}c@{}}Engaged \\ (participant)\end{tabular}} & \textbf{\begin{tabular}[c]{@{}c@{}}Engaged\\ (caregiver)\end{tabular}} & \textbf{\begin{tabular}[c]{@{}c@{}}Disengaged\\ (participant)\end{tabular}} & \textbf{\begin{tabular}[c]{@{}c@{}}Disengaged\\ (caregiver)\end{tabular}} \\ \hline
\textbf{P5}                                                & 5519                 & 1103                    & 1457                                                                      & 683                                                                 & 235                                                                         & 130                                                                    \\
\textbf{P7}                                                & 7028                 & 6327                    & 596                                                                       & 2056                                                                & 161                                                                         & 660                                                                    \\
\textbf{P9}                                                & 5982                 & 6689                    & 471                                                                       & 856                                                                 & 684                                                                         & 463                                                                    \\
\textbf{P11}                                               & 9273                 & 4770                    & 2387                                                                      & 883                                                                 & 1618                                                                        & 234                                                                    \\
\textbf{P12}                                               & 5388                 & 3803                    & 1306                                                                      & 1597                                                                & 899                                                                         & 935                                                                    \\
\textbf{P17}                                               & 3181                 & 3655                    & 684                                                                       & 928                                                                 & 482                                                                         & 475                                                                    \\
\textbf{P18}                                               & 8028                 & 4225                    & 1191                                                                      & 1076                                                                & 586                                                                         & 711                                                                    \\ \hline
\end{tabular}%
}
\label{tab:clip-counts}
\end{table}

\section{Methodology}
\label{sec:methods}

\subsection{Diarization Model Evaluation}
To evaluate speaker diarization, two authors independently annotated 14 audio files, representing approximately 20\% of the dataset, using ELAN\footnote{\url{https://archive.mpi.nl/tla/elan}}. Inter-annotator reliability was evaluated on 20\% of this subset using correctness, defined as the proportion of speech correctly labeled, and diarization error rate (DER), which accounts for missed speech, false alarms, and speaker confusion. Using one annotator as reference, the annotations achieved an average DER of 10.0\% and correctness of 96.8\%.

We considered four speaker diarization models: NVIDIA NeMo End-to-End Sortformer \cite{kuchaiev2019nemo, pmlr-v267-park25h}, NVIDIA NeMo Cascaded Clustering Diarizer \cite{kuchaiev2019nemo}, NVIDIA NeMo Cascaded Neural Diarizer \cite{kuchaiev2019nemo, nvidia2023diarmsddtelephonic}, and Pyannote \cite{bredin2023pyannote}. Because speaker overlap was frequent, we excluded the NeMo Cascaded Clustering Diarizer as it did not explicitly predict overlapping speech. We also excluded the NeMo End-to-End Sortformer because it did not adequately distinguish child vocalizations from the robot's. We therefore evaluated Pyannote and the NeMo Cascaded Neural Diarizer in detail.

Both models were evaluated using original and denoised audio;  denoising was performed using the model described by Défossez et al. \cite{defossez2020real}. We evaluated automatic and predetermined speaker-count configurations and compared all configurations using DER. For the NeMo Cascaded Neural Diarizer, we additionally evaluated combinations of voice activity detection and speaker embedding models.

The best-performing Pyannote configuration achieved a mean DER of $29.9 \pm 9.3\%$, compared with $50.9 \pm 10.9\%$ for the best-performing NeMo Cascaded Neural Diarizer configuration. Pyannote produced significantly lower DER, $t(13) = -5.31$, $p < .001$, and was selected for subsequent analyses\footnote{We specifically used Pyannote 3.1 through the HuggingFace: \url{https://huggingface.co/pyannote/speaker-diarization-3.1}}. The selected configuration used original audio and automatically estimated the number of speakers. DER was calculated using non-overlapping speech to assess speaker identification and separation without the uncertainty associated with overlapping segments. This configuration eliminated the need for manual speaker counting and additional denoising, avoiding potential alterations to acoustic features.

Despite naturalistic background noise and frequent speaker overlap, the selected Pyannote configuration achieved a mean DER of $29.9 \pm 9.3\%$. This result was comparable to the DIHARD III Track 2 baseline DER of $19.37$--$28.10\%$ for diarization from raw audio \cite{dihard3results}. Xu et al. \cite{xu2024exploring} reported a DER of $34.6\%$ for Pyannote on child-adult interactions involving children with ASD, excluding overlapping speech, which decreased to $26.4\%$ after fine-tuning. The DER obtained in this work, therefore, fell between their pretrained and fine-tuned Pyannote results on a related child-adult interaction dataset.

\begin{table}[!t]
\caption{Comparison of diarization models shown with 95\% confidence intervals.}
\centering
\label{tab:diarization_comparison}
\footnotesize
\setlength{\tabcolsep}{3pt}
\begin{tabular}{@{}lcccc@{}}
\toprule
\textbf{Model} & 
\textbf{DER (\%)} & 
\textbf{FA (\%)} & 
\textbf{MS (\%)} & 
\textbf{SC (\%)} \\
\midrule
\makecell[l]{NeMo Neural} &
\makecell{$50.98$$\pm 10.89$} &
\makecell{$22.36$$\pm 6.79$} &
\makecell{$12.31$$\pm 4.46$} &
\makecell{$16.30$$\pm 5.65$} \\[2pt]

Pyannote &
\makecell{$29.99$$\pm 9.31$} &
\makecell{$6.69$$\pm 2.68$} &
\makecell{$4.76$$\pm 3.33$} &
\makecell{$18.54$$\pm 7.11$} \\
\bottomrule
\end{tabular}
\end{table}

\subsection{Data Preprocessing}
\begin{figure}[!t]
\centering
\includegraphics[width=1.0\columnwidth]{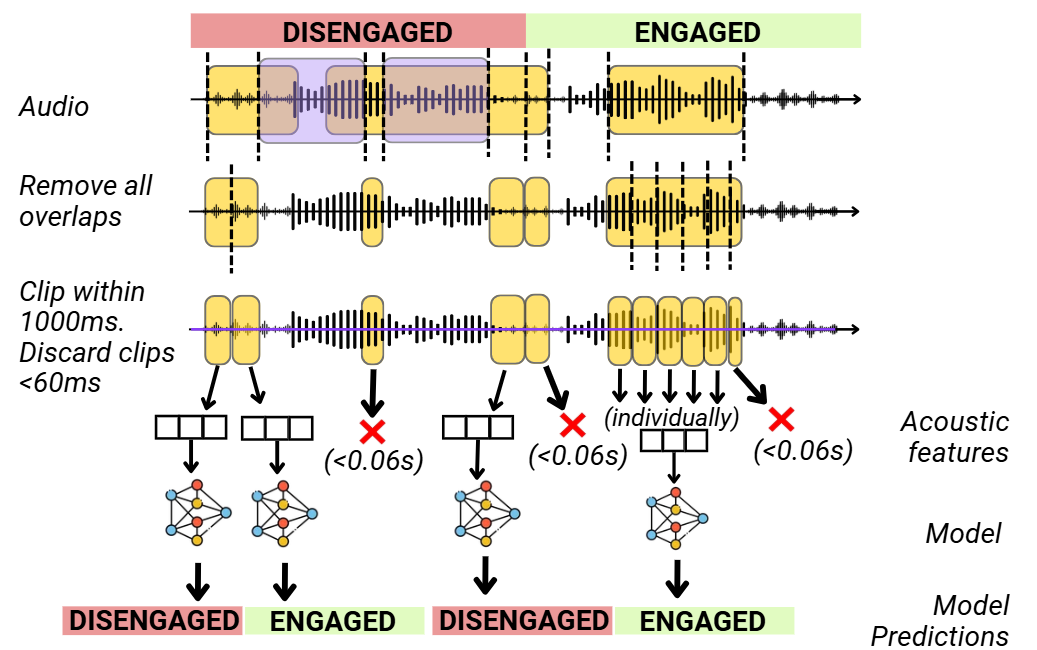}
\caption{Audio pre-processing, handling overlapping speaker segments, multiple speakers within an engagement window, and clipping of audio segments.}
\label{fig:diarization_pipeline}
\end{figure}
Following speaker diarization, we segmented the continuous recordings into speaker-specific audio clips using the temporal boundaries and speaker labels produced by Pyannote, as shown in Figure \ref{fig:diarization_pipeline}. Overlapping speech was excluded to retain only non-overlapping audio for each speaker This resulted in a substantial decrease of available audio data. Although this is a disadvantage of isolating speaker-specific acoustic features in naturalistic environments, it was necessary in order to get acoustic features specific to only one speaker. We identified the speaker labels corresponding to the child participant and caregivers involved in the primary activity and retained only their audio. The resulting clips were segmented into one-second intervals following Jain et al. \cite{jain2020modeling}, and segments shorter than 60 ms were discarded \cite{eyben2015gemaps}. Clip statistics are reported in Table \ref{tab:clip-counts}.

We extracted acoustic features from the speaker-specific clips and corresponding undiarized audio using the extended Geneva Minimalistic Acoustic Parameter Set (eGeMAPS), which comprises 88 features related to acoustic properties, including pitch, voice quality, and loudness \cite{eyben2015gemaps}. eGeMAPS also includes cepstral parameters relevant to modeling affective states \cite{eyben2015gemaps}. An additional speaker id variable was added along with the eGeMAPs features to distinguish between features from child and caregiver audio. The duration of the clip was also added as input to the models.


\subsection{Engagement Modeling and Evaluation}
\label{sec:methods-evaluation}
We trained seven different machine learning models using their default hyperparameters (following Jain et al. \cite{jain2020modeling}): neural network with one hidden layer with 100 nodes (NN), k-nearest neighbors, gradient boosted decision trees (DTs), random decision tree forest, naive Bayes, support vector machine (SVM), and logistic regression. We accounted for class imbalance (Table \ref{tab:clip-counts}) by weighing engaged/disengaged classes by the inverse of their frequency in order to penalize incorrect model predictions for the minority disengaged class.

We trained the models based on the personalized model evaluation setup proposed by Shi et al.\cite{shi2022toward}. In that setup, evaluation was performed on a held-out session of a target participant, while the training set consisted of the available sessions from all other participants, together with the earlier available sessions of the target participant. This is repeated for all sessions for each participant. Thus, the model is exposed to both population-level patterns learned from other participants and participant-specific patterns learned from the target participant's earlier sessions. This adoption of a session-based model evaluation strategy also accounts for the temporal nature of the multi-session dataset and avoids potential temporal leakage, since models are evaluated on chronologically subsequent sessions rather than randomly partitioned instances within a session. Since the number of sessions, and consequently the number of audio clips, varied across participants, we report weighted averages for all metrics to place more importance on model performance for participants with more data. We report F1 score, accuracy, and area under the receiver operating characteristic curve (AUROC) metrics. AUROC is primarily used to evaluate model performance given the class imbalance (Table \ref{tab:clip-counts}).

\section{Results}
\label{sec:results}

\begin{figure}[!b]
    \centering

    \begin{subfigure}{0.8\columnwidth}
        \centering
        \includegraphics[width=\linewidth]
        {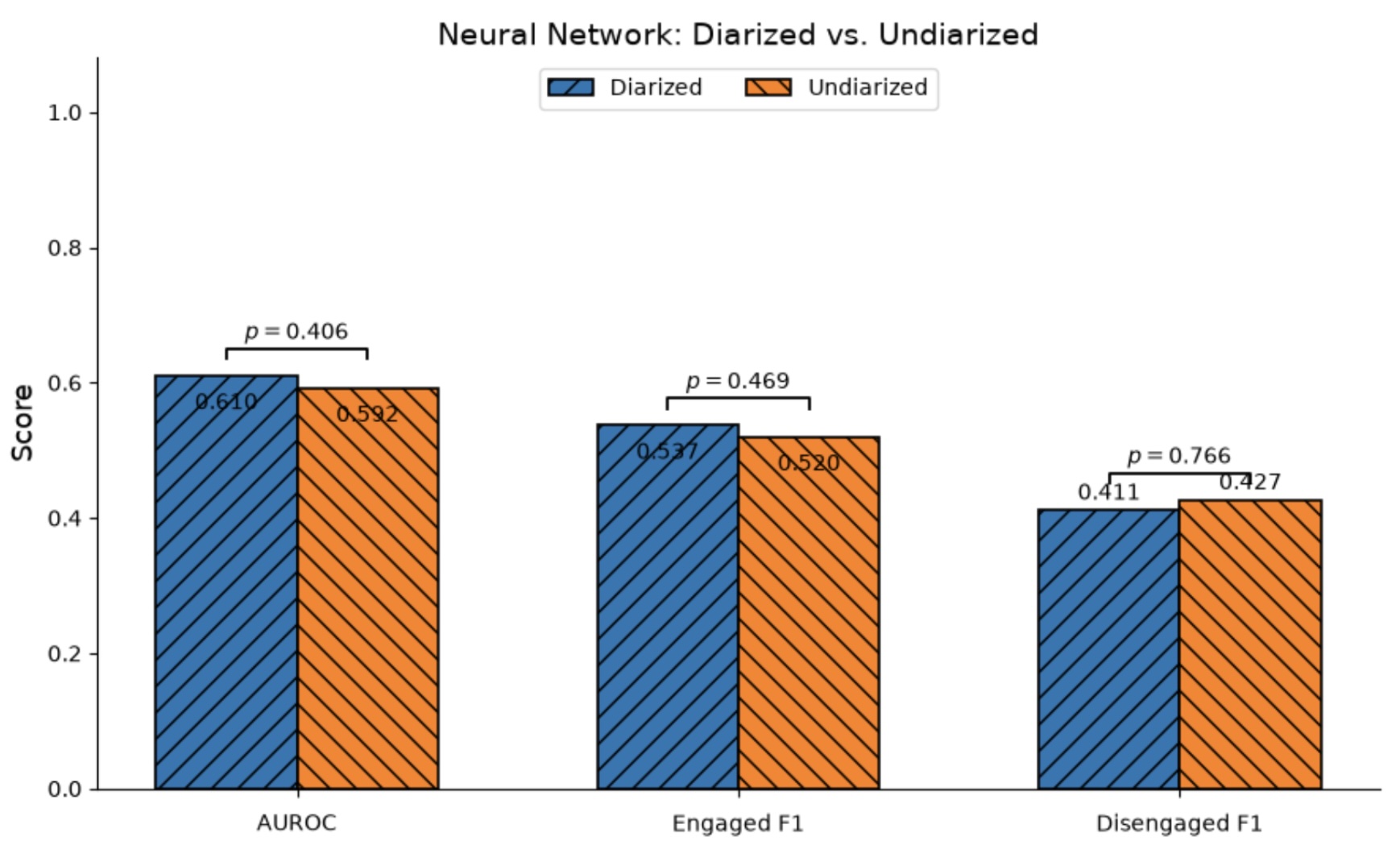}
        \caption{Neural network}
        \label{fig:results-gemaps-nn}
    \end{subfigure}

    \vspace{2mm}

    \begin{subfigure}{0.8\columnwidth}
        \centering
        \includegraphics[width=\linewidth]
        {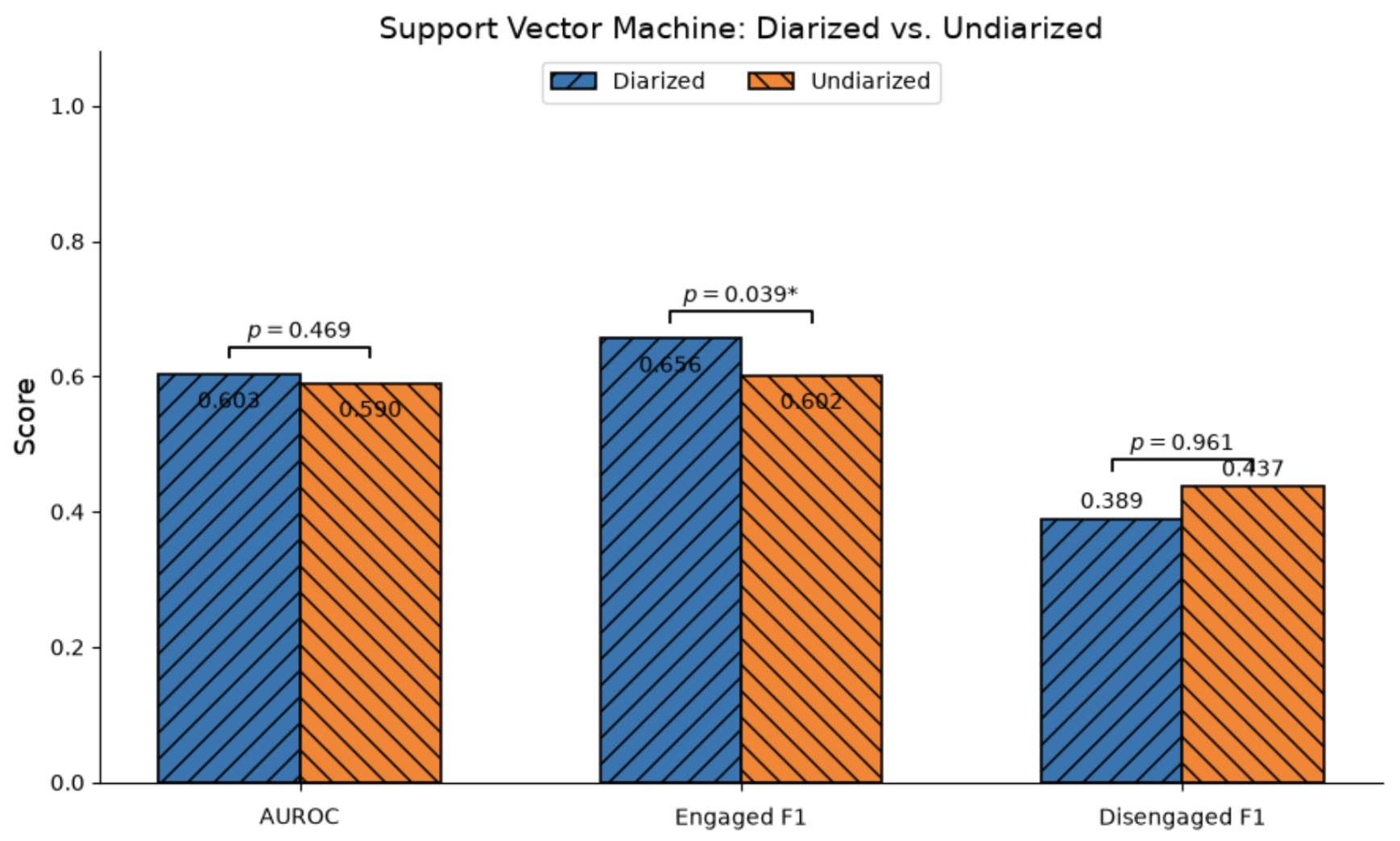}
        \caption{SVM}
        \label{fig:results-gemaps-svm}
    \end{subfigure}

    \vspace{2mm}

    \begin{subfigure}{0.8\columnwidth}
        \centering
        \includegraphics[width=\linewidth]
        {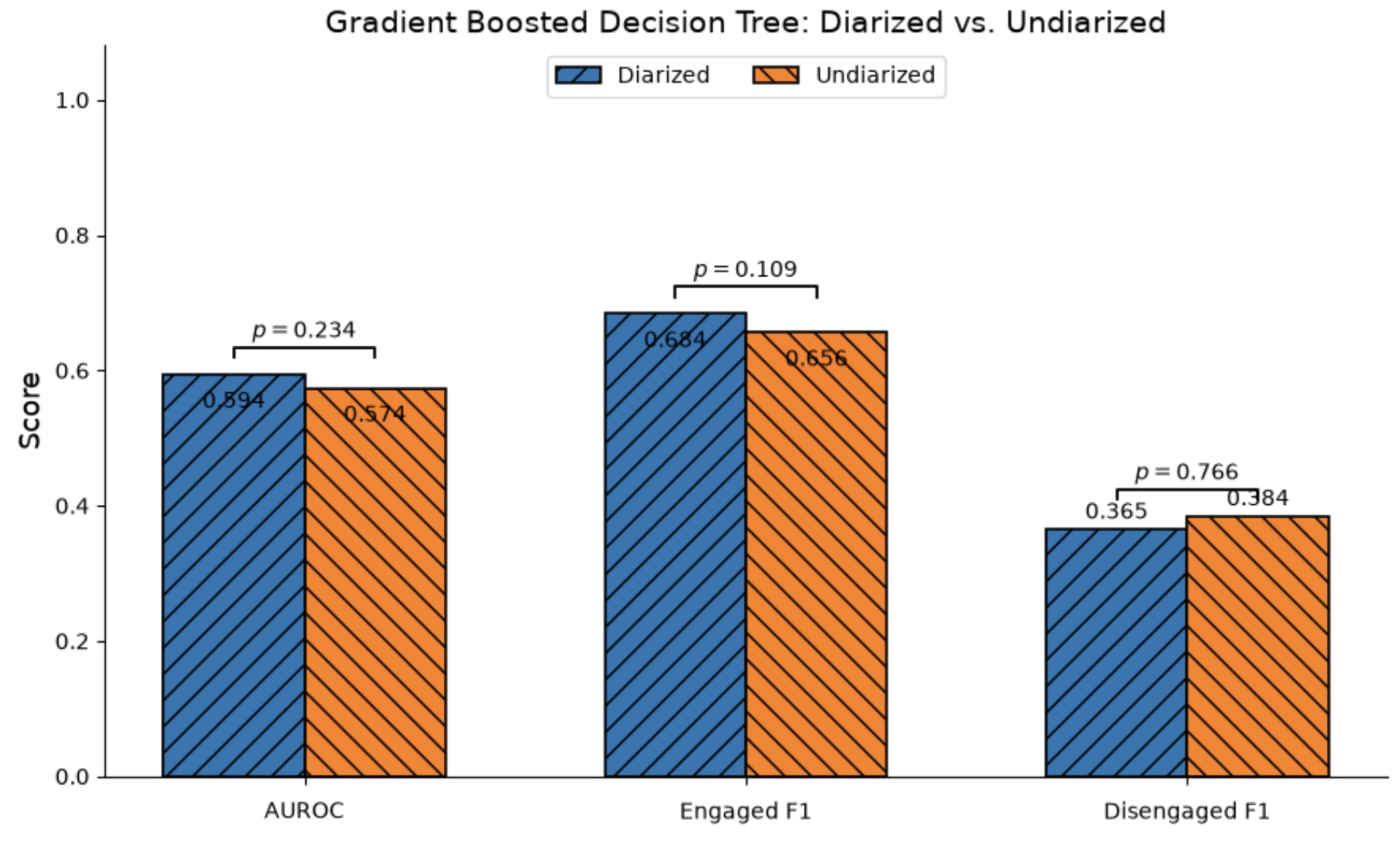}
        \caption{Gradient Boosted Decision Tree}
        \label{fig:results-gemaps-xgboost}
    \end{subfigure}

    \caption{Comparison of three different models trained on diarized versus undiarized eGeMAPs features: (a) neural network, (b) support vector machine, and (c) gradient boosted decision tree.}
    \label{fig:results-gemaps}
\end{figure}

Across the seven evaluated models, we compared performance using diarized and undiarized audio data. We report the three models with the highest AUROC under either condition: NN, SVM, and gradient boosted DT. Across these models, AUROC ranged from 0.594 to 0.610, indicating modest discrimination between engaged and disengaged clips and aligning with prior findings on the limited contribution of eGeMAPS acoustic features to engagement classification \cite{jain2020modeling}. All three models achieved higher AUROC with diarized audio than with undiarized audio (Figure \ref{fig:results-gemaps}); however, a one-sided Wilcoxon signed-rank test indicated that the difference was not statistically significant (NN: $W = 16.00 p = 0.41$, SVM: $W = 15.00, p = 0.47$, gradient boosted DT: $W = 19.0, p = 0.23$). Thus, diarization consistently corresponded with higher discrimination between engagement states, although the statistical analysis did not establish a significant improvement.

AUROC complemented the class-specific F1 scores by measuring discrimination across classification thresholds, which was particularly relevant given the class imbalance in the data (Table \ref{tab:clip-counts}). Engaged-class F1 scores were significantly higher with diarized audio than with undiarized audio for SVM model ($p < .05$), whereas disengaged-class F1 scores did not differ significantly for three models ($p > .05$). The predominance of engaged clips likely contributed to the higher engaged-class F1 scores, while the lower disengaged-class performance aligned with prior findings on the greater variability of disengaged behaviors \cite{jain2020modeling}. Overall, diarization significantly improved classification of the majority engaged class and consistently increased AUROC, although the AUROC improvement was not statistically significant.




\begin{table*}[!t]
\caption{Comparison of model performance using child-speaker and caregiver-speaker audio for all participants.}
\centering
\resizebox{\textwidth}{!}{%
\begin{tabular}{c|cccc|cccc}
\hline
\multirow{2}{*}{\textbf{Participant}} 
& \multicolumn{4}{c|}{\textbf{Child Speaker}} 
& \multicolumn{4}{c}{\textbf{Caregiver Speaker}} \\
& \textbf{AUROC} 
& \textbf{F1 (Engaged)} 
& \textbf{F1 (Disengaged)} 
& \textbf{Accuracy} 
& \textbf{AUROC} 
& \textbf{F1 (Engaged)} 
& \textbf{F1 (Disengaged)} 
& \textbf{Accuracy} \\
\hline

p5  & 0.612 & 0.842 & 0.184 & 0.740 & 0.615 & 0.759 & 0.249 & 0.661 \\
p7  & 0.634 & 0.653 & 0.374 & 0.578 & 0.624 & 0.746 & 0.320 & 0.645 \\
p9  & 0.579 & 0.425 & 0.629 & 0.582 & 0.494 & 0.410 & 0.414 & 0.481 \\
p11 & 0.627 & 0.634 & 0.466 & 0.592 & 0.605 & 0.807 & 0.220 & 0.703 \\
p12 & 0.556 & 0.271 & 0.492 & 0.420 & 0.543 & 0.386 & 0.511 & 0.463 \\
p17 & 0.572 & 0.572 & 0.494 & 0.579 & 0.609 & 0.631 & 0.427 & 0.552 \\
p18 & 0.622 & 0.751 & 0.266 & 0.654 & 0.515 & 0.701 & 0.238 & 0.581 \\

\hline
\end{tabular}%
}
\label{tab:child-parent-results}
\end{table*}

Lastly, for the highest performing model (neural network), we explored the contribution of caregiver-only and child-only acoustic features towards model performance. As shown in Table \ref{tab:child-parent-results}, we see that for five of the child participants (7, 9, 11, 12 and 18), the acoustic features from the child-only speech segments resulted in a higher AUROC compared to caregiver-only speech. For the other two participants (5 and 17), the opposite was the case. Some caregivers were much more involved and prompted the child participants actively, while others were less involved, which may explain why, for some participants, the caregiver's acoustic features were more predictive of participant engagement compared to participant acoustic features, while for others, the participant's acoustic features were more predictive. These results demonstrate that, depending on the participant, the importance of the caregiver's acoustic features can vary. 

\section{Conclusion}
\label{sec:conclusion}
This work examined the effect of speaker diarization on acoustic engagement classification during in-home SAR interventions with children with ASD. Across the highest-performing models, diarized audio consistently produced higher AUROC values than undiarized audio, although the differences were not statistically significant. Diarization significantly improved engaged-class F1 scores for SVM, whereas disengaged-class performance remained low and did not differ significantly between conditions. We further examined child and caregiver acoustic features separately and found that their contributions to engagement classification varied across participants, highlighting individual variability in the acoustic correlates of engagement. Overall, these findings demonstrated the potential value of speaker-specific acoustic information while emphasizing the importance of individual differences for engagement modeling in naturalistic interactions. Overall, these findings extend prior evidence of individual variability in ASD engagement modeling to the acoustic domain and demonstrate the feasibility and potential value of isolating speaker-specific vocalizations for engagement modeling in complex, naturalistic interactions.

\clearpage
\section*{Acknowledgements}
\label{sec:acknowledgments}
Kaleen Shrestha is supported by the National Science Foundation CISE Graduate Fellowships under Grant No. 2313998.

\section*{Compliance with Ethical Standards}
\label{sec:compliance}

The data used in this work was previously collected with approval granted by the University of Southern California Institutional Review Board  (UP-16-00755).

\label{sec:refs}
\balance
\bibliographystyle{IEEEbib}
\bibliography{strings,refs}

\end{document}